\documentclass[prl,twocolumn,aps,floats,showpacs]{revtex4}
\usepackage{graphicx}
\usepackage{psfrag}

\begin{document}

\title{Correspondence between continuous variable and discrete quantum systems of
arbitrary dimensions}

\author{{\v C}aslav Brukner$^1$, Myungshik S. Kim$^2$, Jian-Wei Pan$^1$, Anton Zeilinger$^1$}
\affiliation{$^1$Institut f\"ur Experimentalphysik, Universit\"at Wien, Boltzmanngasse 5, A--1090 Wien, Austria\\
$^2$School of Mathematics and Physics, Queen's University,
Belfast, BT7 1NN, United Kingdom}

\date{\today}

\begin{abstract}

We establish a mapping between a continuous variable (CV) quantum
system and a discrete quantum system of arbitrary dimension. This
opens up the general possibility to perform any quantum
information task with a CV system as if it were a discrete system
of arbitrary dimension. The Einstein-Podolsky-Rosen state is
mapped onto the maximally entangled state in any finite
dimensional Hilbert space and thus can be considered as a
universal resource of entanglement. As an explicit example of the
formalism a two-mode CV entangled state is mapped onto a
two-qutrit entangled state.

\end{abstract}


\maketitle

Quantum information processing enables performance of
communication and computational tasks beyond the limits that are
achievable on the basis of laws of classical physics
\cite{nielsen}. While most of the quantum information protocols
were initially developed for quantum systems with finite
dimensions (qudits) they have also been proposed for the quantum
systems with continuous variables (CV), such as quantum
teleportation \cite{CVteleport}, entanglement swapping
\cite{CVswap}, entanglement purification \cite{CVpurify}, quantum
computation \cite{CVcompute}, quantum error correction
\cite{CVcorrect}, quantum dense coding \cite{CVdense}, and
quantum cloning \cite{CVclone}.

With the exception of two-mode bipartite Gaussian states
\cite{separability} there are no general criteria to test
separability of a general state in infinite-dimensional Hilbert
spaces. Similarly, the demonstration of the violation of Bell's
inequalities for CV systems is based predominantly on the
phase-space formalism \cite{phasespace} and the generalization to
CV systems of various Bell's inequalities derived for discrete
systems and the criteria for their violation remains open. It is
thus highly desirable to find mapping between CV and discrete
systems. This would open up the possibility the CV systems to be
exploited to perform quantum information tasks as if they were
qudits, by applying protocols which are already developed for
discrete $d$ -dimensional systems. It also would allow to apply
all criteria known for discrete systems for the classification of
states (e.g. for separability or for violation of Bell's
inequalities) to CV systems.

Very recently a mapping between CV systems and qubits
(two-dimensional systems) was established \cite{pan,mista}. This
enables to construct a Clauser-Horne-Shimony-Holt (CHSH)
inequality \cite{chsh} for CV systems \cite{pan}, without relying
on the phase-space formalism and to analyze the separability of
the infinite-dimensional Werner states \cite{mista}. It was shown
in Ref. \cite{pan} that the Einstein-Podolsky-Rosen (EPR)
\cite{epr} state
\begin{equation}
|\mbox{EPR}\rangle = \int dq \mbox{ } |q\rangle_1 \otimes |q\rangle_2,
\label{epr}
\end{equation}
where $|q\rangle_1 \otimes |q\rangle_2$ denotes a product state
of two subsystems of a composite system, maximally violates the
CHSH inequality, a question which remained unanswered within the
phase-space formalism. This is important because the EPR state -
the maximally entangled state of CV systems - is considered as a
natural resource of entanglement in CV quantum information
processing.

It is intuitively clear that the potentiality of an
infinite-dimensional system as a resource for quantum information
processing goes beyond that of the qubit system. In particular,
as it will be shown below, the CHSH inequality for CV systems can
be maximally violated even with non-maximally entangled states.
To show the full potential of infinite dimensional systems it
will be important to find a mapping between CV and discrete
quantum systems of arbitrarily high dimensions. An example of the
use of mapping is to check the violation of Bell's inequalities
for higher-dimensional systems \cite{collins} by the EPR state.
Such a mapping is also necessary if one wants to implement those
quantum information tasks developed for discrete systems to CV
systems, which exclusively requires higher-dimensional Hilbert
spaces. These are, for example, the quantum key distribution
based on higher alphabets \cite{crypto} and the quantum solutions
of the coin-flipping problem \cite{coinflip}, of the Byzantine
agreement problem \cite{byzantine}, and of a certain
communication complexity problem \cite{complexity}.

In this paper we establish a mapping between a CV and a discrete
system of arbitrary dimension. Mathematically, for an
infinite-dimensional Hilbert space we construct the generators of
SU(n) algebra for finite $n$, which build up the structure of a
$n$-dimensional Hilbert space. This allows to consider a CV
system as representing a quantum system of any dimension, i.e. a
CV system can be used in various quantum information tasks even
those which require systems of different dimensions. In
particular, the EPR state is always mapped onto the maximally
entangled state in any finite dimensional Hilbert space. Thus it
can be considered as a universal resource of entanglement.

Any Hermitian operator on a $n$-dimensional Hilbert space can be
expanded into the identity operator $\hat{1}_n$ and the
generators of the SU(n) algebra. We use a description which was
introduced in Ref. \cite{hioe} (See also Ref. \cite{mahler}). One
can introduce transition-projection operators
\begin{equation}
\hat{P}_{jk}=|j \rangle\langle k|,
\end{equation}
where $|j\rangle$ with $j=1,...,n$ are orthonormal basis vectors
on the Hilbert space of dimension $n$. The operators
$\hat{P}_{jk}$ will next be used to define another set of
$n^2\!-\!1$ operators, which are formed in three groups and are
denoted by the symbols $\hat{u}$, $\hat{v}$ and $\hat{w}$. One
defines
\begin{eqnarray}
\hat{u}_{jk}\!&=&\!\hat{P}_{jk}+\hat{P}_{kj} \label{masta} \\
\hat{v}_{jk}\!&=&\!i(\hat{P}_{jk}-\hat{P}_{kj}) \\
\hat{w}_l\!&=&\!-\sqrt{\frac{2}{l(l+1)}} (\hat{P}_{11}+\hat{P}_{22}+...+\hat{P}_{ll}-l\hat{P}_{l+1,l+1})
\label{prezir}
\end{eqnarray}
where $1 \!\leq l \!\leq n-1$ and $1\! \leq j \! < k \! \leq n$.

It is easy to check that when $n\!=\!2$, these operators are the
ordinary Pauli (spin) operators along the $x$, $y$ and $z$
direction. In general, the operators $\hat{u}$, $\hat{v}$ and
$\hat{w}$ generate the algebra SU(n). That is, the vector
$\vec{\hat{s}}=(\hat{u}_{12},...,\hat{v}_{12},...,\hat{w}_1,...,\hat{w}_{n-1})$
has components $\hat{s}_j$  $(j=1,...,n^2\!-\!1)$ that satisfy
the algebraic relation
\begin{equation}
[\hat{s}_j,\hat{s}_k]=2i f_{jkl} \hat{s}_l,
\label{vozim}
\end{equation}
where repeated indices are summed from 1 to $n^2-1$, and $f_{jkl}$ is the completely antisymmetric
structure constant of the SU(n) group.

It can be shown that the operators $\hat{s}_j$ fulfill the
relations $\mbox{Tr}(\hat{s}_j)\!=\!0$ and $\mbox{Tr}(\hat{s}_i
\hat{s}_j)\!=\!2\delta_{ij}$. This enables to decompose any
Hermitian operator in a $n$-dimensional Hilbert space as linear
sums of $\hat{s}_j$. To extend the formalism to operators acting
in the Hilbert space of composite systems the direct product of
$\hat{s}_j$ (i.e. $\hat{s}_i \otimes...\otimes \hat{s}_k$) is used
for a basis. Then the general quantum state $\hat{\rho}$ of a
composite system consisting of $L$ systems with dimension $n$ and
observable $\hat{a}$ which can be measured on such a system can be
represented by \cite{hioe,mahler}
\begin{eqnarray}
\hat{\rho} &=&\sum_{x_1,...,x_L=0}^{n^2-1} t_{x_1...x_L}
\hat{s}_{x_1}
\otimes ... \otimes \hat{s}_{x_L} \label{state} \\
\hat{a} &=& \sum_{x_1,...,x_L=1}^{n^2-1} a_{x_1...x_L}
\hat{s}_{x_1} \otimes ... \otimes \hat{s}_{x_L}, \label{operator}
\end{eqnarray}
respectively, where $\hat{s}_0\!=\!\hat{1}_n$. The vector with
components $t_{x_1...x_L}$ is the generalized Bloch vector, which
is real due to the hermiticity of $\hat{\rho}$. Specifically
$t_{0...0}\!=\!1/n^L$ (so that $Tr(\hat{\rho})=1$) and
$t_{x_1...x_L}\!=\!1/2^L \mbox{Tr}(\hat{\rho} \hat{s}_{x_1}
\otimes ... \otimes \hat{s}_{x_L})$ for $x_1,...,x_L \in
\{1,...,n^2-1\}$. The expectation value of the observable
$\hat{a}$ in the state $\hat{\rho}$ is given by
\begin{equation}
\mbox{Tr}(\hat{\rho}\hat{a})=2^L \sum_{x_1,...,x_L=1}^{n^2-1}
t_{x_1...x_L} a_{x_1...x_L}
\end{equation}

We now establish an algebraic equivalence between Hilbert spaces of different dimensionality.
For a given Hilbert space of dimension $N$
we first construct the generators of SU(n) algebra for $n \leq N$,
which build up the structure of a $n$-dimensional Hilbert space.
In the limit $N\rightarrow \infty$ we then obtain a mapping
between a CV system and a discrete system of dimension $n$.

We introduce the transition-projection operators
\begin{equation}
\hat{P}_{jk}(m)=| n m + j \rangle\langle n m + k|,
\label{denis}
\end{equation}
where $0 \!\leq m \!\leq \!\left[\frac{N}{n}\right]\!-\!1$ and
$1\! \leq \!j \!< \!k \!\leq \!n$. Here $\left[\frac{N}{n}\right]$
denotes the integer part of $\frac{N}{n}$. For each $m$ one
constructs the $n^{2}\!-\!1$ operators
\begin{eqnarray}
\hat{u}_{jk}(m)&=&\hat{P}_{jk}(m)+\hat{P}_{kj}(m) \label{ris}\\
\hat{v}_{jk}(m)&=&i(\hat{P}_{jk}(m)-\hat{P}_{kj}(m)) \\
\hat{w}_l(m)&=&-\sqrt{\frac{2}{l(l+1)}} (\hat{P}_{11}(m)+\hat{P}_{22}(m)+... \nonumber \\
&+& \hat{P}_{ll}(m)-l\hat{P}_{l+1,l+1}(m)),\label{sir}
\end{eqnarray}
where $1\!\leq l \!\leq n-1$. Thus the initial Hilbert space of
dimension $N$ is divided into a series of subspaces of dimension
$n$. Within each such subspace (indexed by $m$) the set of
operators $\vec{\hat{s}}(m)=$ $(\hat{u}_{12}(m),...$
$,\hat{v}_{12}(m),...,$ $\hat{w}_1(m),...,\hat{w}_{n-1}(m))$ are
defined according to Eqs. (\ref{ris}-\ref{sir}). They are
generators of the SU(n) algebra because they satisfy the
algebraic relation (\ref{vozim}) by the definition.

Next, we define the operators
\begin{eqnarray}
\hat{U}_{jk}&=&\oplus\sum_{m=0}^{[N/n]} \hat{u}_{jk}(m), \label{skakavac} \\
\hat{V}_{jk}&=&\oplus\sum_{m=0}^{[N/n]} \hat{v}_{jk}(m), \label{joj} \\
\hat{W}_{jk}&=&\oplus\sum_{m=0}^{[N/n]} \hat{w}_{jk}(m), \label{zaso}
\end{eqnarray}
where $\oplus$ denotes the direct sum of operators. The central
point in the construction of the mapping is the introduction of
the set of operators $\vec{\hat{S}}=\oplus\sum_{m=0}^{[N/n]}
\vec{\hat{s}}(m)= (\hat{U}_{12},..., \hat{V}_{12},...,
\hat{W}_1,...,\hat{W}_{n-1})$. This set has elements $\hat{S}_j$'s
$(j=1,...,n^2\!-\!1)$ which also satisfy the general algebraic
relation (\ref{vozim}). This can easily be proved as follows
\begin{eqnarray}
[\hat{S}_j,\hat{S}_k] &=&[\oplus\sum_{m} \hat{s}_j(m), \oplus\sum_{r} \hat{s}_k(r)]=
\oplus\sum_{m,r} [\hat{s}_j(m),\hat{s}_k(r)] \nonumber \\
&=& \oplus\sum_{m} 2i f_{jkl} \hat{s}_l(m) = 2i f_{jkl} \hat{S}_l.
\label{trcim}
\end{eqnarray}
Note that $[\hat{s}_j(m),\hat{s}_k(r)]\!=\!0$ if $m\! \neq\! r$.
Therefore the set of operators $\vec{\hat{S}}$ generate the SU(n)
algebra as well. However, in contrast to the set of generators
$\vec{\hat{s}}(m)$ which acts on $n$-dimensional subspaces, the
set $\vec{\hat{S}}$ acts on the full $N$-dimensional Hilbert
space. It can be shown that for $n\!=\!2$ the three SU(2)
operators are the "pseudospin" operators introduced in Ref.
\cite{pan,mista}.

So far we have built up the structure of a $n$-dimensional
Hilbert space from the original Hilbert space of a higher
dimension $N$. Note, that the SU(n) generators as given by Eq.
(\ref{skakavac}-\ref{zaso}) can be defined for all $n \leq N$.
However only if $N$ is exactly divisible by $n$ all $N$
dimensions of the original Hilbert space are exploited; otherwise
less than $N$. In what follows we use this algebraic equivalence
to establish a concrete correspondence between quantum states and
observables of two systems, one with dimension $n$ and one with
dimension $N$, with $N>n$.

Note that here $\mbox{Tr} \hat{S}_i\!=\!0$ and
$\mbox{Tr}(\hat{S}_i\hat{S}_j)\!=\!2[N/n]\delta_{ij}$. With any
operator $\hat{a}$ (as given in Eq. (8)) acting in a Hilbert
space of $L$ $n$-dimensional systems, we associate the operator
\begin{equation}
\hat{A} = \sum_{x_1,...,x_L=1}^{n^2-1} a_{x_1...x_L} \hat{S}_{x_1}
\otimes ... \otimes \hat{S}_{x_L},
\label{mrtvo}
\end{equation}
in a Hilbert space of $L$ $N$-dimensional systems, with the
coefficients $a_{x_1...x_n}$ which are the same as in the
decomposition (8) of $\hat{a}$. This establishes a correspondence
between the full set of observables in a $n$-dimensional Hilbert
space with a specific subset of observables in a $N$-dimensional
Hilbert space.

From the physical perspective two quantum systems can be
considered as equivalent if the probabilities for outcomes of all
possible future experiments performed on one and on the other
system are the same. This suggests to establish a correspondence
between the quantum states of the two Hilbert spaces as follows.
With any state $\hat{\rho}$ (as given in Eq. (9)) of $L$
$n$-dimensional systems we associate a class $[\Omega]$ of states
of $L$ $N$-dimensional systems with the property that the
expectation value of any observable $\hat{a}$ measured in
$\hat{\rho}$ is equal to the expectation value of the observable
$\hat{A}$ measured in every of the states from the class
$[\hat{\Omega}]$. Mathematically, the mapping is established by
the requirement
$\mbox{Tr}(\hat{\rho}\hat{a})=\mbox{Tr}(\hat{\Omega} \hat{A})$
for any $\hat{a}$ and associated $\hat{A}$ and for any state
$\hat{\Omega}$ from the class $[\hat{\Omega}]$. If the
measurements are constrained to the type (\ref{mrtvo}), the proper
expectation value $\mbox{Tr}(\hat{\Omega} \hat{A})$ can be
obtained if one represents the class $[\hat{\Omega}]$
mathematically by $[\hat{\Omega}]\!:=\!
\sum_{x_1,...,x_L=1}^{n^2-1} T_{x_1...x_L} \hat{S}_{x_1} \otimes
... \otimes \hat{S}_{x_L}$ with
$T_{x_1...x_L}\!=\!\frac{1}{[N/n]^L} t_{x_1...x_L}$.

Taking the limit $N\!\rightarrow\! \infty$ for
$\mbox{Tr}(\hat{\Omega} \hat{A})$ one obtains the mapping between
an expectation value measured on a CV system, and the expectation
value measured on a discrete system of arbitrary dimension. Note
that only expectation values (probabilities) have operational
meaning. To give an example of different infinite-dimensional
states that belong to the same class $[\Omega]$ consider the
maximally entangled state $ |\psi\rangle \!= \!\lim_{N
\rightarrow \infty} \frac{1}{\sqrt{N}} \sum_{i=0}^{N-1}
|j\rangle_1 \otimes |j\rangle_2$ and the mixture $\hat{w}\!=\!
\oplus \sum_{m=0}^{\infty} p(m) |\psi(m)\rangle \langle \psi(m)|$
(with $\sum_{m=0}^{\infty} p(m)\!=\!1)$ of maximally entangled
states $|\psi(m)\rangle \!=\! \frac{1}{\sqrt{n}} \sum_{j=0}^{n-1}
|n m +j \rangle_1 \otimes | n m +j\rangle_2,$ in different
$n\times n$-dimensional subspaces of the original Hilbert space.
Both of them are mapped onto the maximally entangled state
$|\psi\rangle \!=\!\frac{1}{\sqrt{n}} \sum_{i=0}^{n-1}
|j\rangle_1 \otimes |j\rangle_2$ in an $n \times n$-dimensional
space. This example shows that even non-maximally entangled
states can be considered as a resource of maximal entanglement in
lower dimensional Hilbert spaces. For example, the mixture
$\hat{w}$ introduced above for $n\!=\!2$ can maximally violate
the CHSH inequality of Ref. \cite{pan}.

However it is important to note that the EPR state is the only
state which is mapped onto the maximally entangled state in any
finite dimensional Hilbert space. Thus the violation of Bell's
inequalities for arbitrarily high dimensional systems
\cite{collins} or various quantum protocols which use maximally
entangled states of different dimensions
\cite{crypto,byzantine,complexity} can all be demonstrated by the
EPR state.

Experimentally, a state produced by nondegenerate optical
parametric amplifier (NOPA state) can be considered as the
"regularized" EPR state (note that the original EPR state
(\ref{epr}) is unnormalized) \cite{Banaszek}. The NOPA state is
given by
\begin{equation}
|\mbox{NOPA}\rangle=\sum_{k=0}^{\infty} \frac{(\tanh r)^k}{\cosh r} |k\rangle_1 \otimes |k\rangle_2
\label{crta}
\end{equation}
where $r > 0$ is the squeezing parameter and $|k\rangle_1 \otimes
|k\rangle_2$ is a product of the Fock states of the two modes. It
becomes the optical analog of the EPR state in the limit of high
squeezing \cite{Banaszek}.

To give an explicit example for the application of our method we
will map the NOPA state onto an entangled state of two qutrits.
This is important if one wants to use the NOPA state in quantum
communication protocols which are developed for systems of two
entangled qutrits (see \cite{complexity}). We will analyze the
violation by the NOPA state of the Bell inequality for two
qutrits \cite{kaszlikowski,collins}. The Bell inequality is given
as $B\!\leq\! 2$, where $B$ (the Bell expression) is a certain
combination of probabilities for the measurements of two qubits
and 2 is the limit imposed by local realistic models. In Ref.
\cite{fu} the violation of the Bell inequality is investigated
for the states of the form $|\psi\rangle = \sum_{k=0}^{2} a(k)
|k\rangle_1 \otimes |k\rangle_2$ and for a restricted class of
observables which are constructed by unbiased symmetric
beam-splitters \cite{marek}. Here $a(k)$ are real coefficients and
$|k\rangle_1 \otimes |k\rangle_2$ are orthonormal basis states of
two qutrits. The maximal value for the Bell expression was found
to be $B_{max}\!=\!4 |a(1) a(2)| \!+\!
4/\sqrt{3}(|a(1)a(3)|\!+\!|a(2)a(3)|)$ (if $a(1) \!\geq \!a(2)
\!\geq \!a(3)$ and $\mbox{max}\{a(1),a(2),a(3)\}\! \leq\!
\sqrt{6+3\sqrt{3}}/2$, which is our case of study).

The Bell expression in quantum mechanics is given by the
expectation value of a certain operator (the Bell operator). The
general method for establishing our correspondence between CV and
discrete systems dictates that the expectation value of the Bell
operator in a two-qutrit state is equal to the expectation of the
associated Bell operator measured on the NOPA state. This implies
that the entangled two-qutrit state onto which the NOPA is mapped
is of the form as given above with the coefficients $a(k)= (\tanh
r)^k/\sqrt{1+(\tanh r)^2+(\tanh r)^4}$, $k \in \{0,1,2\}$. These
states can be obtained by projecting the NOPA state onto any of
the $3 \times 3$-dimensional subspaces spanned by the states $|n
m +j \rangle_1 \otimes | n m +k\rangle_2$ for a given $m$ and
$j,k \in \{0,1,2\}$.

The amount of violation of the Bell inequality as a function of
the squeezing parameter $r$ is given in Fig. 1a and 1b for
different ranges of $r$. Interestingly, in the interval $r\in
[0,0.5]$ there is no violation. This explicitly shows that for
the set of observables considered in
\cite{kaszlikowski,collins,fu} not even all pure entangled states
violate the Bell inequality. Further, the maximal violation
$(B=2.9011)$ is at $r=1.4998$; not for $r\rightarrow \infty$
which one would expect. This again explicitly confirms the more
general result that non-maximally entangled states can violate
Bell's inequality more strongly than the maximally entangled one
\cite{acin}. Finally, the Bell expression for $r \rightarrow
\infty$ reaches asymptotically the value $2.87293$ which is also
the value obtained for the maximally entangled two-qutrit state.
This is understandable as in that limit the NOPA state becomes
the EPR state and thus is mapped onto the maximally entangled
two-qutrit state.

\begin{figure}
\includegraphics[angle=0,width=9cm]{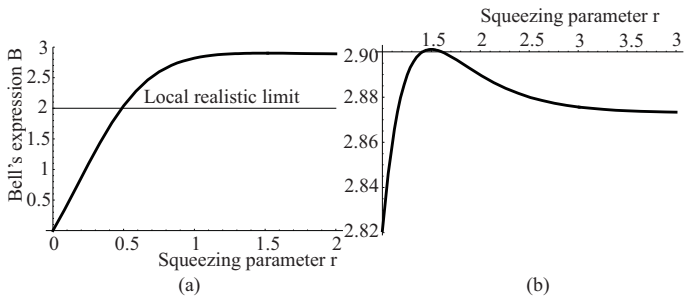}
\caption{The Bell expression $B$ for the NOPA state as a function
of the squeezing parameter $r$ for different ranges of $r$. The
NOPA state is mapped onto a state of two entangled qutrits for
which the Bell inequality for qutrits $B\!>\!2$ is analyzed. In
the interval $[0,0.5]$ of $r$ there is no violation (a). For
$r\!>\!0.5$ the amount of violation of the Bell inequality
increases with an increase of $r$, until it reaches the maximal
value at $r=1.4998$ ($B=2.9011$). With further increase of $r$,
$B$ begins to decrease reaching asymptotically the value of
$2.87293$ (b).} \label{bild}
\end{figure}

In this paper we use the representation in terms of the generators
of the SU(n) algebra to establish the correspondence between CV
and discrete systems. The particular representation is of no
importance; other representations could also be possible.
However, the central point should always be the use of the
transition-projector operators as given in Eq. (\ref{denis}).

In conclusion, we find a correspondence between the CV quantum
systems and discrete quantum systems of arbitrary dimension. This
enables to apply {\it all} results of the physics of quantum
information processing known for discrete systems also to CV
systems.

This work is supported by the Austrian  FWF project F1506, and by
the QIPC program of the EU. MSK acknowledges the financial
support by the UK Engineering and Physical Sciences Research
Council through GR/R33304. We thank Jinhyoung Lee and Wonmin Son
for helpful comments and discussions.

\end{document}